\documentclass[12pt]{article}

\begin{document}

\vspace*{1.0cm}

\begin{center}
{\large 
{\bf Relativistic Equation of State of Nuclear Matter \\
     for Supernova Explosion \\ }}
\vspace*{0.5cm}
H. Shen$^{a,b,}$\footnote{e-mail: shen@rcnp.osaka-u.ac.jp},
H. Toki$^{a,c,}$\footnote{e-mail: toki@rcnp.osaka-u.ac.jp},
K. Oyamatsu$^{c,d,}$\footnote{e-mail: oyak@postman.riken.go.jp}
and K. Sumiyoshi$^{c,}$\footnote{e-mail: sumi@postman.riken.go.jp} \\
\end{center}
      $^{a}$Research Center for Nuclear Physics (RCNP), Osaka University,
           Ibaraki, Osaka 567-0047, Japan \\
      $^{b}$Department of Physics, Nankai University, Tianjin 300071, China \\
      $^{c}$The Institute of Physical and Chemical Research (RIKEN),
            Wako, Saitama 351-0198, Japan \\
      $^{d}$Department of Energy Engineering and Science, Nagoya University,
            Nagoya 464-8603, Japan \\

\vspace*{0.5cm}

\begin{abstract}
	 We construct the equation of state (EOS) of nuclear matter 
at finite temperature and density with various proton fractions 
within the relativistic mean field (RMF) theory for the use in the 
supernova simulations.  We consider nucleus, alpha-particle, proton 
and neutron in equilibrium at densities smaller than about 
$\rho_B \sim 10^{14.2} g/cm^3$ by minimizing the free energy 
of the nuclear matter. The calculation is based on the RMF theory 
with the parameter set, TM1, which has been demonstrated to provide 
good accounts of the ground and the excited state properties of finite 
nuclei.  We tabulate the outcome for various densities as the pressure, 
the free energy, entropy etc, at enough mesh points in the density 
range $ \rho_B = 10^{5.1} \sim 10^{15.4} g/cm^3$ and the temperature range 
$ T=0 \sim 100 MeV$ and the proton fraction range of $ Y_p = 0 \sim 0.56$ 
to be used for the supernova simulations. 
\end{abstract}

\newpage

\section{Introduction}

 The fate of massive stars with mass of more than 8 times the solar mass 
is spectacular in the end when they have used up their nuclear energy by 
creating the Fe core, which has the largest binding energy per particle.  
When the Fe core mass exceeds the Chandrasekhar mass supported by the 
degenerate 
electron pressure against the gravitational force, the Fe core starts to 
collapse until the central density becomes around that of the normal 
matter density.
There the strong repulsive force and the Pauli effect among nucleons work 
against the gravitational collapse and swing back the matter motion outwards. 
If this outward motion is strong enough to blow off the outer layers, 
the stellar explosion is observed as supernova.  This is called the prompt 
explosion.  The strength of the outward shock motion depends crucially 
on the equation of state (EOS) of nuclear matter.

	The general belief in the recent years is, however, that the outward 
shock propagation alone is not strong enough to succeed supernova explosion.  
Most of the gravitational collapse energy is carried away by neutrinos.  
If some small fraction of the neutrino energy is dropped in the stellar 
matter just behind the slowing down shock region with possible help of the 
hydrodynamic instability, the outward shock motion can survive to blow off 
the outer materials.  This is called the delayed explosion.  
Whichever the mechanism of the supernova explosion is, we need detailed 
knowledges on the EOS of nuclear matter and also the neutrino reaction 
rates in nuclear matter in the wide ranges of density and temperature.  
We have started a project to provide these informations, the EOS and 
the neutrino reaction rates, by studying carefully the existing knowledge 
on finite nuclei including the unstable nuclei and the weak processes.  
In this paper, we discuss the results on the EOS.  

	The knowledge on nuclear matter necessary for supernova simulations 
is much more than that obtained from the study of finite nuclei.  
The density may go up to 5 to 10 times the normal matter density in the 
stellar phenomena.  The temperature may climb up to 30 to 50 MeV.  
The proton fraction may go down to zero, the pure neutron matter.  
Hence, we have to make a large extrapolation to all directions from 
our knowledge on nuclear physics which are available from laboratory 
experiments.  Hence, we have to construct a theoretical model, 
which has a strong basis from the microscopic theory and at the same time 
is easy to handle for many observables.  This theoretical model has to 
be checked with many experimental facts on finite nuclei.

	In recent years, there was a very important achievement on 
the theoretical understanding on nuclear physics.  The use of the 
relativistic many body theory enabled us to remove the long standing 
problem on the nuclear saturation property, which is not reproduced 
in the non-relativistic approach \cite{BM90}.  
The relativistic theory can reproduce many observables on the nuclear 
structure including the unstable nuclei and also the nuclear reactions, 
in particular the spin observables \cite{BT92,ST94,SST97,HS97}.  
The relativistic Brueckner-Hartree-Fock (RBHF) theory reproduced 
the saturation property of the nuclear matter very close to 
the 'experimental' value and removed the disease of the non-relativistic 
Brueckner-Hartree-Fock theory, which provided the saturation points 
lying on the Coester line.  The relativistic framework contains very 
strong density dependent repulsive component due to Z-graph three body 
process in the language of non-relativistic framework.  
The strong spin-orbit force and the spin dynamics are the natural 
consequences of the relativistic equation of motion.

	Hence, we chose the relativistic mean field (RMF) theory, 
which is relatively easy to use and at the same time carries the essential 
features of the microscopic theory as the RBHF theory, as a phenomenological 
model to describe nuclei and nuclear matter at various circumstances 
\cite{HS97}.  
First, the RMF lagrangian was constructed to provide the results of the RBHF, 
for which we include the non-linear $\omega$ meson coupling term in addition 
to the generally used non-linear $\sigma$ meson coupling terms.  
We fixed the model parameters as the coupling constants and the masses 
so as to reproduce many key experimental data on the ground state masses 
and the radii of nuclei including unstable ones \cite{ST94}.  
The parameter set was 
named TM1 and then applied to many observables as deformed nuclei, 
triaxial deformations, nuclear excitations \cite{HTT95,HS96,MT97}.  
The agreement with the existing data is extremely satisfactory.  
Hence, the RMF theory with the TM1 parameter set could go through 
the critical check from Nuclear Physics.  
The RBHF results for the symmetric nuclear matter were also 
compared very nicely up to high density.

	This was the place we decided to go on to construct the EOS table 
for the astrophysical use.  We want to state here our philosophy.  
The purpose of making this table with only the nuclear degree of freedom 
is to work out the best EOS with the known experimental and the theoretical 
knowledge on nuclear physics. This EOS should be considered as the basis EOS.
At high density and temperature, however, 
there may be the need of the mixture of hyperons, various mesons and even 
quarks.  As our knowledges on these degrees of freedom become more solid, 
we should include them for the EOS table, in particular at high density.  

	We work out the EOS of nuclear matter in the wide parameter ranges 
in density, temperature and proton fraction.  The density range is 
$\rho_B = 10^{5.1} \sim 10^{15.4} g/cm^3$,  the temperature range is 
$T= 0 \sim 100 MeV$ and the proton fraction is $Y_p = 0 \sim 0.56$.  
In very high density and high temperature region, the nuclear matter is 
homogeneous and hence the calculation is not difficult.  However, when 
the density is below $\rho_B \sim 1/3 \rho_0$ the matter becomes 
inhomogeneous.  In order to treat the inhomogeneous matter, 
we take the Thomas-Fermi approximation, where the inhomogeneous matter 
is composed of a lattice of spherical nuclei with protons, neutrons and 
alpha-particles \cite{O93}.  
We then minimize the free energy of the nuclear system with respect to 
the proton, neutron and alpha-particle densities in the nuclear
interior and the exterior regions.  This procedure has to be performed 
for each temperature, density and the proton fraction, which is extremely 
cumbersome.  At the same time, we have to repeat the minimization 
procedures with good accuracy in order to get the smooth physical quantities 
as the free energy, pressure and the entropy for the use of supernova 
simulations.  Some part of the present research project is published 
in Ref.\cite{STOS98}.  The published materials do not contain 
the contributions of the alpha-particles.  The new part of this paper 
is the inclusion of the alpha-particles and hence we shall make detailed 
comparisons of the outcome with respect to the alpha-particle contributions.

	This paper is arranged as follows.  We describe the RMF theory 
and the TM1 parameter set in Sect.2 and the formalism of the Thomas-Fermi 
approximation with the alpha-particles in Sect.3.  The numerical results 
are discussed in Sect.4.  The conclusion of this paper is stated in Sect.5. 
We describe the definitions of the physical quantities
tabulated in the EOS table in Appendix A.

\section{ Relativistic mean field theory}

We adopt the RMF theory with non-linear $\sigma$ and $\omega$ terms
to describe homogeneous nuclear matter \cite{ST94}.
We start with the lagrangian given by
\begin{eqnarray}
{\cal L}_{RMF} & = & \bar{\psi}\left[i\gamma_{\mu}\partial^{\mu} -M
-g_{\sigma}\sigma-g_{\omega}\gamma_{\mu}\omega^{\mu}
-g_{\rho}\gamma_{\mu}\tau_a\rho^{a\mu}
\right]\psi  \\ \nonumber
 && +\frac{1}{2}\partial_{\mu}\sigma\partial^{\mu}\sigma
-\frac{1}{2}m^2_{\sigma}\sigma^2-\frac{1}{3}g_{2}\sigma^{3}
-\frac{1}{4}g_{3}\sigma^{4} \\ \nonumber
 && -\frac{1}{4}W_{\mu\nu}W^{\mu\nu}
+\frac{1}{2}m^2_{\omega}\omega_{\mu}\omega^{\mu}
+\frac{1}{4}c_{3}\left(\omega_{\mu}\omega^{\mu}\right)^2   \\ \nonumber
 && -\frac{1}{4}R^a_{\mu\nu}R^{a\mu\nu}
+\frac{1}{2}m^2_{\rho}\rho^a_{\mu}\rho^{a\mu} .
\end{eqnarray}
Here, $\psi$ denotes a SU(2) baryon field of mass $M$ (proton and neutron);
$\sigma, \omega^{\mu},$ and $\rho^{a\mu}$ are 
$\sigma, \omega,$ and $\rho$ meson fields with masses 
$m_{\sigma}, m_{\omega},$ and $m_{\rho}$, respectively.
$W^{\mu\nu}$ and $R^{a\mu\nu}$ are the antisymmetric field tensors 
for $\omega^{\mu}$ and  $\rho^{a\mu}$, which can be written as
\begin{eqnarray}
W^{\mu\nu} & = & \partial^{\mu}\omega^{\nu}
               - \partial^{\nu}\omega^{\mu} ,     \\ 
R^{a\mu\nu} & = & \partial^{\mu}\rho^{a\nu}- \partial^{\nu}\rho^{a\mu}
 +g_{\rho}\epsilon^{abc}\rho^{b\mu}\rho^{c\nu} .
\end{eqnarray}
In the lagrangian, the constants $g_{\sigma}, g_{\omega}, g_{\rho}$ 
are the coupling constants
for the interactions between mesons and nucleons, the coefficients $g_2$
and $g_3$ are the self-coupling constants for $\sigma$ meson field, 
while $c_3$ is the self-coupling constant for $\omega$ meson field.
It is known that the inclusion of the non-linear $\sigma$ terms is
essential to reproduce the properties of nuclei quantitatively and
a reasonable value for the incompressibility, while the
non-linear $\omega$ term is added in order to reproduce the density
dependence of the vector part of the self-energy of the nucleon obtained in
the RBHF theory. 
The lagrangian contains the meson masses, the coupling constants, and the
self-coupling constants as parameters. 
We adopt the parameter set TM1 listed in Table 1,
which was determined in Ref.\cite{ST94} as the best one to reproduce
the properties of finite nuclei in the wide mass range in the periodic
table including neutron-rich nuclei.
The RMF theory with the TM1 parameter set
was also shown to reproduce satisfactory agreement with experimental data
in the studies of the nuclei with deformed configuration \cite{HTT95} and the
giant resonances within the RPA formalism \cite{MT97}.
With the TM1 parameter set, the symmetry energy is 36.9 MeV and
the incompressibility is 281 MeV.

Starting with the lagrangian, we derive a set of the Euler-Lagrange 
equations. The Dirac equation for the nucleon field is given by 
\begin{equation}
\left(i\gamma_{\mu}\partial^{\mu} -M
-g_{\sigma}\sigma-g_{\omega}\gamma_{\mu}\omega^{\mu}
-g_{\rho}\gamma_{\mu}\tau_a\rho^{a\mu}
\right)\psi=0 ,
\end{equation}
and the Klein-Gordon equations for the meson fields are given by
\begin{eqnarray}
\partial_{\nu}\partial^{\nu}\sigma + m_{\sigma}^2\sigma & =
 &-g_{\sigma}\bar{\psi}\psi
 -g_{2}\sigma^{2}-g_{3}\sigma^{3},  \\  
\partial_{\nu}W^{\nu\mu} +m_{\omega}^2\omega^{\mu}  & =
 & g_{\omega}\bar{\psi}\gamma^{\mu}\psi
 -c_{3}\left(\omega_{\nu}\omega^{\nu}\right)\omega^{\mu}, \\  
\partial_{\nu}R^{a\nu\mu} +m_{\rho}^2\rho^{a\mu}     & =
 & g_{\rho}\bar{\psi}\tau_a\gamma^{\mu}\psi
 +g_{\rho}\epsilon_{abc}\rho^{b}_{\nu}R^{c\nu\mu} .
\end{eqnarray}
These equations are coupled nonlinear quantum field equations, which
are very difficult to solve exactly. We adopt the relativistic
mean field approximation as described in Ref.\cite{SW86}, 
in which the meson fields are treated as classical fields,
and the field operators $\sigma, \omega^{\mu},$ and $\rho^{a\mu}$ 
are replaced by their expectation values 
$\langle\sigma\rangle, \langle\omega^{\mu}\rangle,$ 
and $\langle\rho^{a\mu}\rangle$.
In the present study, we consider first static infinite matter so that
we obtain simplified equations, where the derivative terms
in the Klein-Gordon equations vanish automatically 
due to the translational invariance of infinite matter.
The spatial components of the vector meson fields vanish under the rotational
symmetry. For the isovector-vector meson field $\rho^{a\mu}$, 
only the third isospin 
component has a nonvanishing value because of the charge conservation.
Hence, the equations for meson fields are reduced to 
\begin{eqnarray}
\sigma_0 & \equiv & \langle\sigma\rangle =
 -\frac{g_{\sigma}}{m_{\sigma}^2} \langle\bar{\psi}\psi\rangle
 -\frac{1}{m_{\sigma}^2}\left(g_{2}\sigma_0^{2}+g_{3}\sigma_0^{3}\right)
,  \\  
\omega_0 & \equiv & \langle\omega^0\rangle =
 \frac{g_{\omega}}{m_{\omega}^2} \langle\bar{\psi}\gamma^0\psi\rangle
 -\frac{1}{m_{\omega}^2}c_3\omega_0^3
,  \\  
\rho_0   & \equiv & \langle\rho^{30}\rangle =
 \frac{g_{\rho}}{m_{\rho}^2} \langle\bar{\psi}\tau_3\gamma^0\psi\rangle .
\end{eqnarray}
The Dirac equation then becomes 
\begin{equation}
\left(-i\alpha_k\nabla^k + \beta M^{*}
+g_{\omega}\omega_0 +g_{\rho}\tau_3\rho_0 \right)\psi_{is}=
\varepsilon_{is} \psi_{is} ,
\end{equation}
where the index $i$ denotes the isospin degree of freedom (proton and neutron)
and $s$ denotes the index of eigenstates of nucleon,
$M^{*} \equiv M+g_{\sigma}\sigma_0$ is the effective mass,
$\varepsilon_{is}$ is the single-particle energy.

Nucleons occupy single-particle orbits with the occupation probability 
$f_{is}$. At zero temperature, $f_{is}=1$ under Fermi surface,
while $f_{is}=0$ above Fermi surface.
For finite temperature, the occupation probability is given by Fermi-Dirac
distribution,
\begin{eqnarray}
 f_{is}=\frac{1}{1+\exp\left[\left(\varepsilon_{is}-\mu_{i}\right)/T\right]}
       =\frac{1}{1+\exp\left[\left(\sqrt{k^2+{M^{*}}^2}-\nu_{i}\right)
        /T\right]}
,  \\  
f_{\bar{i}s}=\frac{1}{1+\exp\left[\left(-\varepsilon_{\bar{i}s}+\mu_{i}
       \right)/T\right]}
 =\frac{1}{1+\exp\left[\left(\sqrt{k^2+{M^{*}}^2}+\nu_{i}\right)/T\right]} .
\end{eqnarray}
Here the index $i$ and $\bar{i}$ denote nucleons and antinucleons, 
$\varepsilon_{is}$ and $\varepsilon_{\bar{i}s}$ are the single-particle
energy of nucleons and antinucleons, respectively.
The relation between the chemical potential $\mu_i$ and the kinetic
part of the chemical potential $\nu_i$ is give by
\begin{equation}
 \mu_{i}=\nu_i+g_{\omega}\omega_0 +g_{\rho}\tau_3\rho_0.
\end{equation}
The chemical potential $\mu_i$ is related to the number density of nucleon
$n_i$ as 
\begin{equation}
 n_{i}=\frac{\gamma}{2\pi^2}
       \int_0^{\infty} dk\,k^2\,\left(f_{ik}-f_{\bar{i}k}\right),
\end{equation}
where $\gamma$ is the degeneracy factor for the spin degree of freedom 
($\gamma=2$),
the quantum number $s$ is replaced by the momentum $k$ when we do the
integration in the momentum space instead of the summation of eigenstates.
The equations of meson fields can be written as 
\begin{eqnarray}
\sigma_0 &=& -\frac{g_{\sigma}}{m_{\sigma}^2} \displaystyle{\sum_i 
   \frac{\gamma}{2\pi^2} \int_0^{\infty} dk\,k^2\, 
   \frac{M^{*}}{\sqrt{k^2+{M^*}^2}}
   \left(f_{ik}+f_{\bar{i}k}\right) }
  -\frac{1}{m_{\sigma}^2}\left(g_{2}\sigma_0^{2}+g_{3}\sigma_0^{3}\right)
,  \\  
\omega_0 &=&
   \frac{g_{\omega}}{m_{\omega}^2} 
   \left(n_p+n_n\right)
  -\frac{c_3}{m_{\omega}^2}\omega_0^3
,  \\  
\rho_0   &=&
   \frac{g_{\rho}}{m_{\rho}^2} 
   \left(n_p-n_n\right) .
\end{eqnarray}
Here, $n_p$ and $n_n$ are the proton number density and neutron number density
as defined in Eq.(15), and we denote $n_B=n_p+n_n$ as the baryon number 
density. We solve these equations self-consistently.
The thermodynamical quantities are described 
in Ref.\cite{SW86}, and we just write the expressions here.
The energy density of nuclear matter is given by 
\begin{eqnarray}
\epsilon &=& \displaystyle{\sum_i \frac{\gamma}{2\pi^2} 
   \int_0^{\infty} dk\,k^2\, 
   \sqrt{k^2+{M^*}^2}  \left(f_{ik}+f_{\bar{i}k}\right) }
  +\frac{1}{2}m_{\sigma}^2\sigma_0^2+\frac{1}{3}g_{2}\sigma_0^{3}
  +\frac{1}{4}g_{3}\sigma_0^{4}  \\ \nonumber
 & & +g_{\omega}\omega_0 \left(n_p+n_n\right)
  -\frac{1}{2}m_{\omega}^2\omega_0^2-\frac{1}{4}c_{3}\omega_0^{4}
  \\ \nonumber
 & & +g_{\rho}\rho_0 \left(n_p-n_n\right)
  -\frac{1}{2}m_{\rho}^2\rho_0^2,
\end{eqnarray}
the pressure of nuclear matter is given by 
\begin{eqnarray}
 p &=& \displaystyle{\sum_i \frac{\gamma}{6\pi^2} \int_0^{\infty} dk\,k^2\, 
   \frac{k^2}{\sqrt{k^2+{M^*}^2}}  \left(f_{ik}+f_{\bar{i}k}\right) }
  -\frac{1}{2}m_{\sigma}^2\sigma_0^2-\frac{1}{3}g_{2}\sigma_0^{3}
  -\frac{1}{4}g_{3}\sigma_0^{4}  \\ \nonumber
 & & +\frac{1}{2}m_{\omega}^2\omega_0^2+\frac{1}{4}c_{3}\omega_0^{4}
  +\frac{1}{2}m_{\rho}^2\rho_0^2,
\end{eqnarray}
and the entropy density is calculated by
\begin{equation}
s=\displaystyle{\sum_i \frac{\gamma}{2\pi^2} \int_0^{\infty} dk\,k^2\,
\left[-f_{ik}\ln f_{ik}-\left(1-f_{ik}\right)\ln \left(1-f_{ik}\right)
-f_{\bar{i}k}\ln f_{\bar{i}k}-\left(1-f_{\bar{i}k}\right)
\ln \left(1-f_{\bar{i}k}\right) \right] }.
\end{equation}

\section{Thomas-Fermi approximation}

In the range, $T < 15 MeV$ and $\rho_B < 10^{14.2}g/cm^3$, where heavy nuclei 
may be formed in order to lower the free energy, 
we perform the Thomas-Fermi calculation based on the work done 
by Oyamatsu \cite{O93}. In this case, the non-uniform matter can be modeled
as a mixture of free neutrons, free protons, alpha-particles, and a single 
species of heavy nuclei, while the leptons can be treated as uniform 
non-interacting particles separately.
For the system with fixed proton fraction, the leptons play no role in the
minimization of the free energy. Hence we mainly
pay attention to baryon contribution in this calculation.
Hereafter, we will not mention the leptons frequently, while we should
keep in mind that there exists an uniform lepton gas existing everywhere.

We assume that each heavy spherical nucleus is
located in the center of a charge-neutral cell consisting of a
vapor of neutrons, protons and alpha-particles. The nuclei form the
body-centered-cubic (BCC) lattice to minimize the Coulomb lattice energy.
It is useful to introduce the Wigner-Seitz cell to simplify the energy of
the unit cell. The Wigner-Seitz cell is a sphere whose volume is
the same as the unit cell in the BCC lattice.
We define the lattice constant $a$ as the cubic
root of the cell volume. Then, we have
\begin{equation}
V_{cell}=a^3=N_B / n_B,
\end{equation}
where $N_B$ and $n_{B}$ are the
baryon number per cell and the average baryon number density, respectively.
We calculate the Coulomb energy using this Wigner-Seitz approximation and
add a correction energy for the BCC lattice \cite{O93}. This correction
energy is negligible unless the nuclear size is comparable to the cell size.

We assume the nucleon distribution functions $n_i(r)$
($i=n$ for neutron, $i=p$ for proton)
in the Wigner-Seitz cell as
\begin{equation}
n_i\left(r\right)=\left\{
\begin{array}{ll}
\left(n_i^{in}-n_i^{out}\right) \left[1-\left(\frac{r}{R_i}\right)^{t_i}
\right]^3 +n_i^{out},  & 0 \leq r \leq R_i \\
n_i^{out},  & R_i \leq r \leq R_{cell} \\
\end{array} \right. ,
\end{equation}
where $r$ represents the distance from the center of the nucleus and
$R_{cell}$ is the radius of the Wigner-Seitz cell defined by the relation,
\begin{equation}
V_{cell} \equiv \frac{4 \pi}{3} R_{cell}^3.
\end{equation}
The density parameters
$n_i^{in}$ and $n_i^{out}$ are
the densities at $r=0$ and $ r \geq R_i $, respectively. The parameters
$R_i$ and $t_i$ determine the boundary and the relative surface thickness
of the heavy nucleus.

For the distribution function of alpha-particles $n_{\alpha}(r)$,
which should decrease as $r$ approaches the center of the heavy nucleus, 
we assume
\begin{equation}
n_{\alpha}\left(r\right)=\left\{
\begin{array}{ll}
-n_{\alpha}^{out} \left[1-\left(\frac{r}{R_p}\right)^{t_p}
\right]^3 +n_{\alpha}^{out},  & 0 \leq r \leq R_p \\
n_{\alpha}^{out},  & R_{p} \leq r \leq R_{cell} \\
\end{array} \right. ,
\end{equation}
which will give $n_{\alpha} (r=0) =0$ 
and $n_{\alpha} (r>R_p) =n_{\alpha}^{out}$.
Here we use the same parameters $R_p$ and $t_p$ for both proton distribution
function and alpha-particle distribution function in order to avoid 
too many parameters in the minimization procedure. 
The parameters $R_n$ and $t_n$ may be a little different from
$R_p$ and $t_p$ due to the additional neutrons forming a neutron skin in the
surface region.
In order to illustrate these parameters, we show the distributions 
of nucleons and alpha-particles in Fig.1 for the case at
$T=10 MeV$, $Y_p=0.3$, and $\rho_B=10^{13.5}g/cm^3$.
Here we define the baryon mass density as $\rho_B=m_u n_B$ with
$m_u$ being the atomic mass unit.
For the system with fixed temperature $T$, proton fraction $Y_p$,
and baryon mass density $\rho_B$,  there are eight independent parameters
in the ten variables; $a, n_n^{in}, n_n^{out}, R_n, t_n,
n_p^{in}, n_p^{out}, R_p, t_p, n_{\alpha}^{out}$.
The thermodynamically favorable state is the one that minimizes
the free energy density with respect to these eight independent parameters.

In this model the free energy density contributed by baryons is given as
\begin{equation}
f\,=\,F_{cell}\,/\,a^3  
=\,\left(\,E_{cell}\,-\,T\,S_{cell}\,\right)\,/\,a^3   ,
\end{equation}
where the free energy per cell $F_{cell}$ can be written as
\begin{equation}
F_{cell}=(E_{bulk}+E_s+E_C)- T S_{cell}=F_{bulk}+E_s+E_C.
\end{equation}
The bulk energy $E_{bulk}$, entropy $S_{cell}$, and bulk free energy
$F_{bulk}$ are calculated by
\begin{eqnarray}
E_{bulk} &=&\int_{cell} \epsilon \left( \, n_n\left(r\right),
\, n_p\left(r\right), \, n_{\alpha}\left(r\right) \, \right) d^3r, \\
S_{cell} &=&\int_{cell} s \left(\, n_n\left(r\right),
\, n_p\left(r\right), \, n_{\alpha}\left(r\right) \, \right) d^3r, \\
F_{bulk} &=&\int_{cell} f \left( \, n_n\left(r\right),
\, n_p\left(r\right), \, n_{\alpha}\left(r\right) \, \right) d^3r.
\end{eqnarray}
Here $\epsilon \left( \, n_n\left(r\right),
\, n_p\left(r\right), \, n_{\alpha}\left(r\right) \, \right)$, 
$s \left( \, n_n\left(r\right),
\, n_p\left(r\right), \, n_{\alpha}\left(r\right) \, \right)$, 
and $f \left( \, n_n\left(r\right),
\, n_p\left(r\right), \, n_{\alpha}\left(r\right) \, \right)$
are the local energy density, entropy density, and free energy density
at the radius $r$, where it can be considered as a mixed uniform matter
of neutrons, protons, and alpha-particles.
The local free energy density is given as the sum of the nucleon 
and alpha-particle contributions;
\begin{equation}
f \left( \, n_n\left(r\right),
\, n_p\left(r\right), \, n_{\alpha}\left(r\right) \, \right)
=f_{N}+f_{\alpha},
\end{equation}
where $f_N$ and $f_{\alpha}$ are the contributions from nucleons and 
alpha-particles, respectively.
We use the RMF theory to calculate the free energies of neutrons and protons
when density is not extremely low ($n_n+n_p>10^{-5}fm^{-3}$), 
while the classical ideal gas approximation is adopted at low density
($n_n+n_p<10^{-5}fm^{-3}$). The connection 
between the RMF results and the ideal gas approximation is very smooth.

The alpha-particles are treated as non-interacting Boltzman particles,
which is considered as a reasonable approximation because the number 
density of alpha-particles is not so large even though the average
baryon density is high. The free energy density of alpha-particles 
is given by
\begin{equation}
f_{\alpha} =-T\,n_{\alpha}\left[\ln (8n_Q/n_{\alpha})+1\right]
           -n_{\alpha} B_{\alpha}, 
\end{equation}
where we have used the abbreviation 
$n_Q=\left(\frac{M\,T}{2\pi\hbar^2}\right)^{3/2}$,
$B_{\alpha}=28.3 MeV$ is the binding energy of an alpha-particle
relative to free neutrons taken from Ref.\cite{LS91}.
We have to take into account the volume of alpha-particle, otherwise the 
alpha-particle fraction will become some big number at high
densities, where the alpha-particles should actually disappear.
After we take into account the volume excluded by the alpha-particles,
the free energy densities of nucleons and alpha-particles become,
\begin{eqnarray}
f_N (n_n, n_p) &=& (1-u) f_N (\tilde{n}_n, \tilde{n}_p), \\ 
f_{\alpha} (n_{\alpha}) &=& (1-u) f_{\alpha} (\tilde{n}_{\alpha}), 
\end{eqnarray}
where $u=n_{\alpha}\,v_{\alpha}$ is the fraction of space occupied by 
alpha-particles, the effective volume of an alpha-particle,
$v_{\alpha}=24 fm^{-3}$, is taken from Ref.\cite{LS91}.
We denote the effective number density of neutrons ($i=n$),
protons ($i=p$), or alpha-particles ($i=\alpha$) as $\tilde{n}_i=n_i/(1-u)$.
The inclusion of the volume excluded by the alpha-particles
has negligible effect in the low density region, 
while it is necessary for the calculation at high density.

As for the surface energy term $E_s$ due to the inhomogeneity of nucleon
distribution, we take a simple form as,
\begin{equation}
E_s=\int_{cell} F_0 \mid \nabla \left( \, n_n\left(r\right)+
    n_p\left(r\right) \, \right) \mid^2 d^3r.
\end{equation}
The parameter $F_0=70 \, MeV\cdot fm^5$ is determined by doing the Thomas-Fermi
calculations of finite nuclei so as to reproduce the gross properties of
nuclear mass, charge radii and the beta stability line as described in
the appendix in Ref.\cite{O93}.

The Coulomb energy per cell $E_C$ is calculated using the Wigner-Seitz 
approximation and add a correction term for the BCC lattice \cite{O93}
\begin{equation}
E_C=\frac{1}{2}\int_{cell} e \left[n_p\left(r\right)
+2n_{\alpha}\left(r\right)-n_e\right]\,\phi(r) d^3r
\,+\,\triangle E_C,
\end{equation}
where $\phi(r)$ stands for the electrostatic potential calculated 
in the Wigner-Seitz approximation,
$n_e$ is the electron number density of uniform electron gas ($n_e=Y_p\,n_B$),
$\triangle E_C$ is the correction term for the BCC lattice,
which can be approximated as
\begin{equation}
\triangle E_C=C_{BCC}\frac{(Z_{non}e)^2} {a},
\end{equation}
here $a$ is the lattice constant as defined in Eq.(22),
the coefficient $C_{BCC}=0.006562$ is taken from Ref.\cite{O93},
$Z_{non}$ is the non-uniform part of the charge number per cell
given by
\begin{equation}
Z_{non}=\int_{0}^{R_p} 
(n_p^{in}-n_p^{out}-2 n_{\alpha}^{out})
\left[1-\left(\frac{r}{R_p}\right)^{t_p}\right]^3
4\pi r^2 dr.
\end{equation}
Because of the long-range nature of the Coulomb interaction,
the Coulomb energy will depend on the lattice type.
This dependence was discussed in more details in Ref.\cite{O93}.
The system prefers the BCC lattice because the BCC lattice 
gives the lowest Coulomb energy.

\section{Equation of state of nuclear matter}

We construct the EOS table of nuclear matter in the wide ranges of the
temperature $T$, proton fraction $Y_p$, and baryon mass density 
$\rho_B$ as listed in Table 2 for the use of supernova simulations, 
and we also add the results for zero temperature case ($T=0$) and 
pure neutron matter case ($Y_p=0$).
The physical quantities we provide in the table and their definitions
are described in Appendix A. The leptons can be considered as uniform 
non-interacting particles, which are relatively easy to deal with,
so we provide the baryon part of EOS without the lepton contributions
in this table.

For each $T$, $Y_p$, and $\rho_B$, we have to determine the thermodynamically
favorable state of nuclear matter, which has the lowest free energy in 
this model. We perform the minimization of the free
energy for both non-uniform matter and uniform matter.
Here the phase of heavy nuclei formed together with free nucleons
and alpha-particles is referred to as non-uniform matter
while the phase of nucleons mixed with alpha-particles without
heavy nuclei is referred to as uniform matter.
For non-uniform matter, the minimization procedure is realized by using
the Thomas-Fermi method, which includes eight independent parameters 
as described in Sect.3.
For uniform matter, we do the minimization with respect to
converting two protons and two neutrons into an alpha-particle,
so there is only one independent parameter in this minimization procedure.
By comparing the free energies of the non-uniform matter and uniform matter,
we determine the most favorable state of nuclear matter at this $T$, $Y_p$, 
and $\rho_B$. 

We discuss first the phase transition between uniform matter and non-uniform
matter. The phase transition at zero temperature has been discussed 
in Ref.\cite{STOS98}, which does not change after including alpha-particle,
because both the proton fraction and the alpha-particle fraction at zero
temperature are equal to zero. 
In Fig.2, we show the phase boundary dependent on the temperature $T$
at $Y_p=0.01, 0.1, 0.2,$ and $0.4$. The solid curve is the boundary
of non-uniform matter phase where the heave nuclei are formed so that
the heavy nucleus fraction is more than zero denoted by $X_A>0$.
The dashed curve is the boundary where the alpha-particle fraction
changes between $X_{\alpha}<10^{-4}$ and $X_{\alpha}>10^{-4}$.
The definitions of $X_A$ and $X_{\alpha}$ are given in Appendix A.
At low density, the thermodynamically favorable state is
homogeneous nucleon gas with a small fraction of alpha-particles, 
the heavy nuclei are formed at some medium densities where the system 
can lower the free energy by forming heavy nuclei, 
then it becomes uniform matter as density increases
beyond $\rho_B \sim 10^{14.2} g/cm^3$. 
It is seen in this figure that the density where the non-uniform matter
phase appears depends on the temperature very strongly, while the 
density where the non-uniform matter phase disappears is nearly 
independent of the temperature.
As temperature increases, the density range of the non-uniform matter phase
becomes narrower. The non-uniform matter phase disappears completely
at high temperature.
In Fig.3, we show the phase boundary dependent on the proton fraction $Y_p$
at $T=0, 1, 5,$ and $10 MeV$, this dependence is relatively weak 
except at very low $Y_p$ limit. The non-uniform matter phase disappears
at lower temperature with a small $Y_p$ because it is difficult to form 
heavy nuclei with a small proton number.

We plot in Fig.4 the fractions of neutrons, protons, alpha-particles, 
and heavy nuclei as functions of the baryon mass density $\rho_B$ 
at $Y_p=0.2$ with various temperatures. The alpha-particle fraction
$X_{\alpha}$ increases as baryon mass density $\rho_B$ increases, 
but the formation of heavy nuclei brings rapid drops for 
the alpha-particle fraction and the nucleon fractions, because the heavy
nuclei use up most of the nucleons. 
From this figure, we can find the alpha-particle fraction
becomes smaller as the temperature increases.

In Figures 5a and 5b, we display the nuclear mass number $A$ and the proton
number $Z$ as defined in Appendix A 
as a function of the baryon mass density $\rho_B$ with various proton
fraction $Y_p$ at $T=1 MeV$. Both $A$ and $Z$ increase sharply 
before the phase transition. The nucleus becomes larger before the 
system turns into the uniform matter phase. This
behavior is similar to the zero temperature case in Ref.\cite{STOS98}.

We treat the uniform matter and non-uniform matter consistently using
the same RMF theory. Hence, all the resulting thermodynamical quantities
are consistent and smooth in the whole range.
Fig.6 shows the free energy per baryon $F$ as a function of the 
baryon mass density $\rho_B$ with various $Y_p$ at $T=0, 1, 5$ and $10 MeV$.
In the low density and finite temperature range,
the free energy is not significantly altered by variation in $Y_p$,
because the interaction between nucleons is very weak at low density.
The inclusion of alpha-particles can reduce the free energy just before 
the formation of heavy nuclei where the alpha-particle fraction 
is some significant number as shown in Fig.4, so that the 
inclusion of alpha-particles can make the phase transition smoother.

The pressure is calculated from $F$ through the following
thermodynamical relation,
\begin{equation}
p  =\left[\,n_B^2 (\partial F/\partial n_B) \,\right]_{T,Y_p}.
\end{equation}
We show in Fig.7 the pressure as a function of $\rho_B$
with various $Y_p$ at $T=0, 1, 5$ and $10 MeV$.
At low density and finite temperature, the pressure is mainly 
provided by the thermal pressure, which is proportional to the
particle number density. When the alpha-particle fraction is a small number,
the pressure is almost equal to $n_B T$ and independent of $Y_p$.
Just before the formation of heavy nuclei, the pressure curve
shows some weak dependence on $Y_p$ due to the increase of $X_{\alpha}$,
then a sharp drop happens due to the formation of heavy nuclei.
For high $Y_p$, the pressure drops to the negative area.
The pressure drop in the non-uniform matter phase has  a strong
$Y_p$ dependence.
In high density region, all curves come back to the positive area,
and have a rapid rise as the density increases.
The behavior of the pressure at high density is determined
predominantly by the contribution of the vector meson $\omega$
and isovector-vector meson $\rho$ as discussed 
in the Refs.\cite{STOS98, SKT95}.

In Fig.8, we show the results of the entropy per baryon as a function of
$\rho_B$ with various $Y_p$ at $T=0.1, 1, 5$ and $10 MeV$.
The effects of the alpha-particles and heavy nuclei are clearly shown in
the case of $T=1 MeV$, the drops around $10^8 g/cm^3$ 
are due to the increase of the alpha-particle fraction,
while the drops around $10^{10} g/cm^3$ 
are due to the formation of heavy nuclei. 
In the case of $T=0.1 MeV$, the non-uniform matter phase appears beyond
the lowest density limit, so the entropy has a strong $Y_p$ dependence
due to the formation of heavy nuclei.
On the other hand, the free gases outside heavy nuclei become 
dominant compositions of nuclear matter at high temperature, then the effect 
of the formation of heavy nuclei is weakened as the temperature increases.

The properties of the EOS for neutron star matter were discussed in details 
in Ref.\cite{STOS98}. The neutron star masses and profiles using this EOS
were shown and compared with other EOS in the same reference.
The properties of the EOS for hot dense matter in supernova core 
including leptons were discussed in Ref.\cite{SKT95}.

\section{Conclusions and discussions}

The relativistic EOS of nuclear matter is designed for the use
of supernova simulations in the wide temperature and density range 
with various proton fractions as listed in Table 2. 
We have adopted the RMF theory with the non-linear $\sigma$ 
and $\omega$ terms, which was demonstrated to be successful in 
describing the properties of both stable and unstable nuclei as well as 
reproducing the self-energies in the RBHF theory. 
We have worked out consistent calculations for uniform matter and non-uniform 
matter within the RMF theory.
The Thomas-Fermi approximation is adopted to describe inhomogeneous 
nuclear matter, which can be modeled as a mixture of free neutrons, 
free protons, alpha-particles, and a single species of heavy nuclei. 
The inclusion of alpha-particles has significant effect around
the formation of heavy nuclei and make the phase transition smoother.

The relativity plays an essential role in describing the nuclear 
saturation and the nuclear structure \cite{BM90}, it also brings 
some distinctive properties in the EOS comparing with the case in 
the non-relativistic framework \cite{STOS98, SKT95}.
Therefore, it is very interesting and important to study the 
astrophysical phenomena such as supernova explosion and the birth
of neutron stars using the relativistic EOS through extensive 
comparison with the previous studies using the non-relativistic EOS.

\section*{Acknowledgment}

H.S. acknowledges the COE program for enabling her to stay at RCNP-Osaka,
where this work was carried out. 
H.S. is grateful for the hospitality during
her 10 months stay at RCNP, where fruitful
discussions are made with the colleagues.
We would like to thank H.Suzuki and
S.Yamada for helpful suggestions and fruitful discussions.
A large portion of the present calculation has been done using 
the supercomputers SX4 at RCNP and VPP500 at RIKEN.

\section*{Appendix A}

We list the definitions of the physical quantities tabulated in the
EOS table

\begin{itemize}
\item 
(0) temperature: $T$ [$MeV$] \\
The temperature $T$ is written at the beginning of each block in the table,
while we use mark 'cccccccc' to divide blocks with different $T$.
\item
(1) logarithm of baryon mass density: $\log_{10}(\rho_B)$   [$g/cm^3$] \\
\item
(2) baryon number density: $n_B$ [$fm^{-3}$] \\
The baryon number density is related to the baryon mass density as
$\rho_B=m_{u} n_B$ with $m_{u}=931.49432 MeV$
being the atomic mass unit taken from Ref.\cite{cons}.
\item 
(3) logarithm of proton fraction: $\log_{10}(Y_p)$  \\
The proton fraction $Y_p$ of uniform matter is defined by
\begin{equation}
Y_p=\frac{n_p+2n_{\alpha}}{n_B}
   =\frac{n_p+2n_{\alpha}}{n_n+n_p+4n_{\alpha}}
\end{equation}
where $n_p$ is the proton number density, $n_n$ is the neutron number density, 
$n_{\alpha}$ is the alpha-particle number density,
and $n_B$ is the baryon number density. 

For non-uniform matter case, $Y_p$ is the average proton fraction defined by 
\begin{equation}
Y_p=\frac{N_p}{N_B}
\end{equation}
where $N_p$ is the proton number per cell, $N_B$ is the baryon number per cell;
\begin{eqnarray}
N_p &=& \int_{cell} \left[\, n_p\left(r\right) + 2n_{\alpha}\left(r\right) 
                      \,\right] d^3r, \\ 
N_B &=& \int_{cell} \left[\, n_n\left(r\right) +
                         n_p\left(r\right) + 4n_{\alpha}\left(r\right) 
                      \,\right] d^3r, 
\end{eqnarray}
$n_p(r)$ and $n_n(r)$ are the proton and neutron density distribution function
given in Eq.(23), 
$n_{\alpha}(r)$ is the alpha-particle density
distribution function given in Eq.(25).
\item 
(4) proton fraction: $Y_p$  \\
\item 
(5) free energy per baryon: $F$  [$MeV$] \\
The free energy per baryon is defined as relative to 
the free nucleon mass $M=938.0 MeV$ in TM1 parameter set as
\begin{equation}
F=\frac{f}{n_B}-M.
\end{equation}
\item 
(6) internal energy per baryon: $E_{int}$  [$MeV$] \\
The internal energy per baryon is defined as relative to 
the atomic mass unit $m_{u}$ as
\begin{equation}
E_{int}=\frac{\epsilon}{n_B}-m_{u}.
\end{equation}
\item 
(7) entropy per baryon: $S$  [$k_B$] \\
The entropy per baryon is related to the entropy density via
\begin{equation}
S=\frac{s}{n_B}.
\end{equation}
\item 
(8) mass number of heavy nucleus: $A$   \\
The mass number of heavy nucleus is defined by
\begin{equation}
A=\int_{0}^{R_A}  \left[\, n_n\left(r\right) + n_p\left(r\right)
  \,\right] 4\pi r^2 dr,
\end{equation}
where $n_n(r)$ and $n_p(r)$ are the density 
distribution functions in the Thomas-Fermi approximation as given in Eq.(23),
$R_{A}$ is the maximum between $R_p$ and $R_n$, which is considered as 
the boundary of the heavy nucleus.
\item 
(9) charge number of heavy nucleus: $Z$   \\
The charge number of heavy nucleus is defined by
\begin{equation}
Z=\int_{0}^{R_A}  \, n_p\left(r\right) \, 4\pi r^2 dr, 
\end{equation}
\item 
(10) effective mass: $M^{*}$  [$MeV$] \\ 
The effective mass is obtained in the RMF theory for uniform matter.
In non-uniform matter case, the effective mass is a function of space 
due to inhomogeneity of nucleon distribution, so it is meaningless 
to list the effective mass for non-uniform matter. 
We replace the effective mass $M^{*}$ by the free nucleon mass $M$ 
in the non-uniform matter phase.
\item 
(11) free neutron fraction: $X_n$  \\
The free neutron fraction is given by
\begin{equation}
X_n=(n_n^{out} V^{out})/(n_B V_{cell})
\end{equation}
where $V_{cell}$ is the cell volume as defined in Eq.(22), 
$V^{out}=V_{cell}-\frac{4\pi}{3}R_{A}^3$ is the volume outside
heavy nucleus, $n_n^{out}$ is the free neutron number density
outside heavy nucleus, while $n_B$ is the average baryon number density.
\item 
(12) free proton fraction: $X_p$  \\
The free proton fraction is given by
\begin{equation}
X_p=(n_p^{out} V^{out})/(n_B V_{cell})
\end{equation}
with $n_p^{out}$ is the free proton number density
outside heavy nucleus.
\item 
(13) alpha-particle fraction: $X_{\alpha}$  \\
The alpha-particle fraction is defined by
\begin{equation}
X_{\alpha}=4N_{\alpha}/(n_B V_{cell})
\end{equation}
where $N_{\alpha}$ is the alpha-particle number per cell obtained by
\begin{equation}
N_{\alpha}=\int_{cell} n_{\alpha}\left(r\right) d^3r,
\end{equation}
$n_{\alpha}(r)$ is the alpha-particle density
distribution function given in Eq.(25).
\item 
(14) heavy nucleus fraction: $X_A$  \\
The heavy nucleus fraction is defined by
\begin{equation}
X_A=A/(n_B V_{cell})
\end{equation}
where $A$ is the mass number of heavy nucleus as defined in Eq.(47).
\item 
(15) pressure: $p$ [$MeV/fm^3$] \\
The pressure is calculated through the following
thermodynamical relation
\begin{equation}
p  =\left[\,n_B^2 (\partial F/\partial n_B) \,\right]_{T,Y_p}
\end{equation}
\item 
(16) chemical potential of neutron: $\mu_n$ [$MeV$] \\
The chemical potential of neutron relative to
the free nucleon mass $M$ is calculated through
the thermodynamical relation
\begin{equation}
\mu_n=\left[\,\partial (n_B F) /\partial n_n  \,\right]_{T,n_p} 
\end{equation}
here $n_n=(1-Y_p)\,n_B$. 
\item 
(17) chemical potential of proton: $\mu_p$ [$MeV$] \\
The chemical potential of proton relative to
the free nucleon mass $M$ is calculated through 
the thermodynamical relation
\begin{equation}
\mu_p=\left[\,\partial (n_B F) /\partial n_p  \,\right]_{T,n_n}
\end{equation}
here $n_p=Y_p\,n_B$. 
\end{itemize}

We have done the following check for the EOS table,\\
(1) the consistency of the thermodynamical quantities,
\begin{equation}
        F + \frac{p}{n_B}=\mu_n (1-Y_p)+\mu_p Y_p,
\end{equation}
(2) the consistency of the fractions,
\begin{equation}
        X_n+X_p+X_{\alpha}+X_A=1,
\end{equation}
(3) the consistency of the relation between $F$, $E_{int}$, and $S$,
\begin{equation}
        F=E_{int}-TS +m_{u} -M.
\end{equation}
Gernerally these consistency relations can be satisfied 
within the accuracy of 0.001.
Physical constants to convert units are taken from Ref.\cite{cons}.




\section*{Figure captions}

\begin{description}
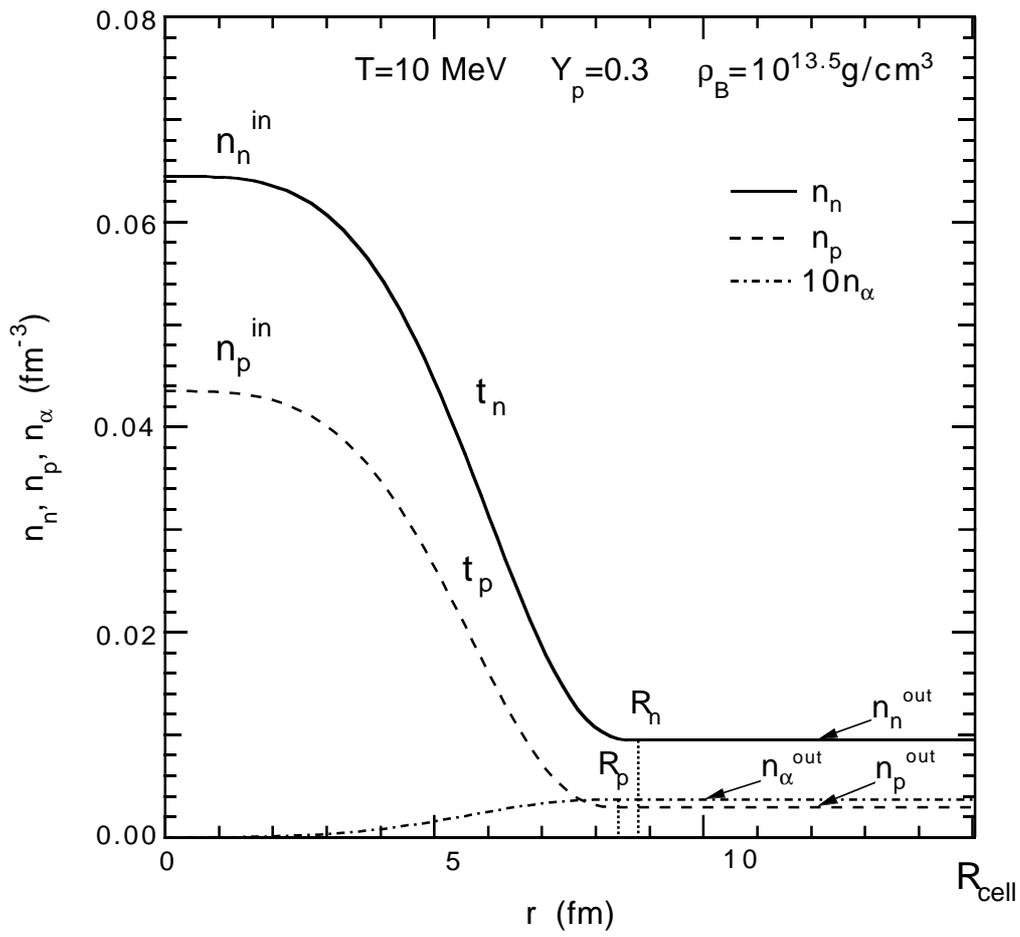
\item[Fig.1.] The distributions of neutrons, protons, and alpha-particles 
in the Wigner-Seitz cell at $T=10 MeV$, $Y_p=0.3$, and $\rho_B=10^{13.5}g/cm^3$
are plotted by solid curve, dashed curve, and dot-dashed curve, respectively.
In order to show the alpha-particle distribution clearly, 
we enlarge it by a factor 10.
The parameters $n_i^{in}, n_i^{out}, R_i$ and $t_i$
of the density distributions are defined in Eqs.(23) and (25),
while $R_{cell}$ represents the radius of
spherical Wigner-Seitz cell.

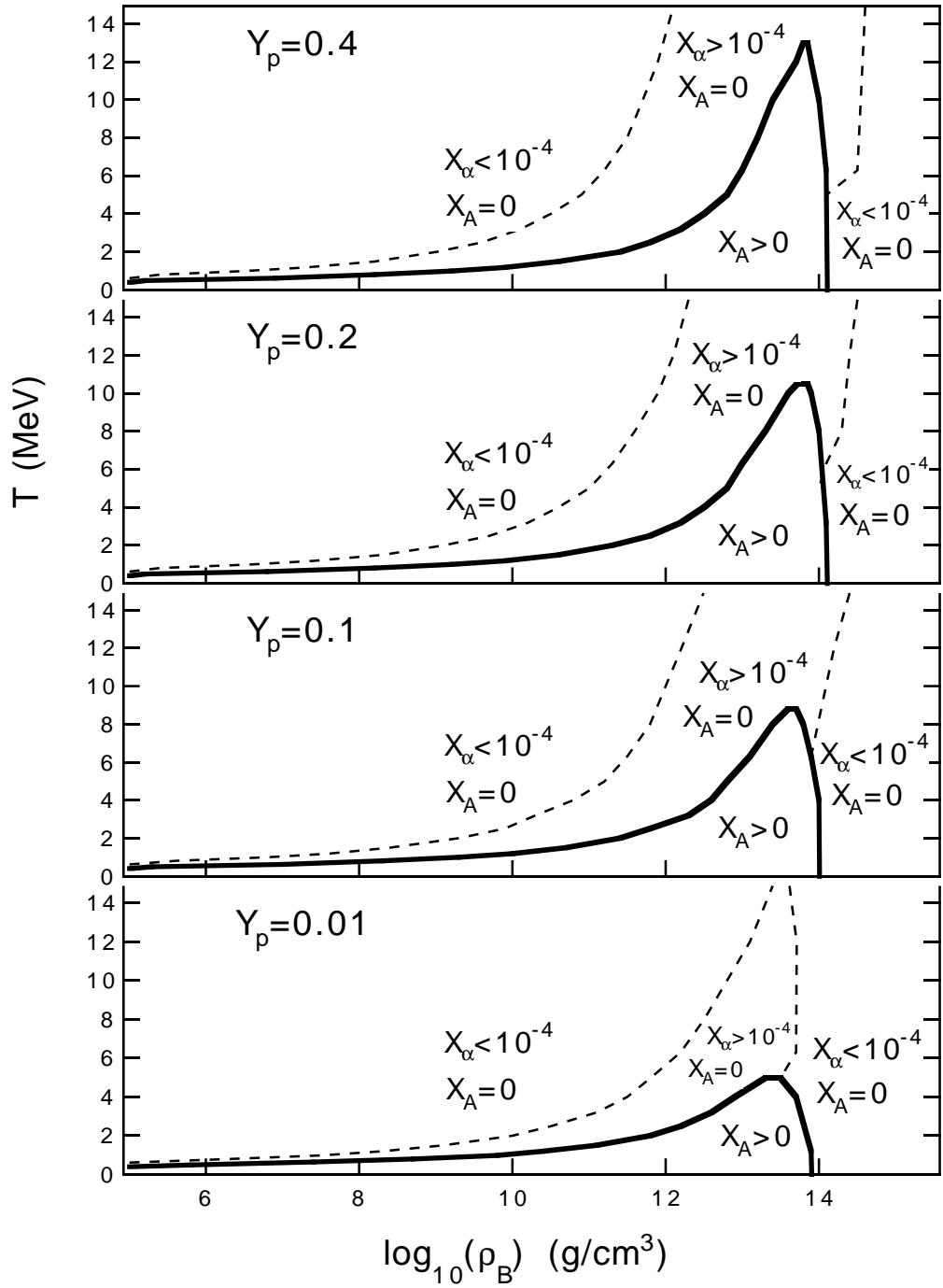
\item[Fig.2.] The phase diagrams of the nuclear matter at
$Y_p=0.01, 0.1, 0.2,$ and $0.4$ in $\rho_B - T$ plane.
$X_{\alpha}$ and $X_A$ are alpha-particle fraction and heavy nucleus
fraction, respectively. The regions denoted by $X_A>0$ correspond
to the non-uniform matter phase. The solid curves show the boundary
between the non-uniform matter phase and uniform matter phase,
while dashed curves show the boundary where the
alpha-particle fraction changes between
$X_{\alpha}<10^{-4}$ and $X_{\alpha}>10^{-4}$.

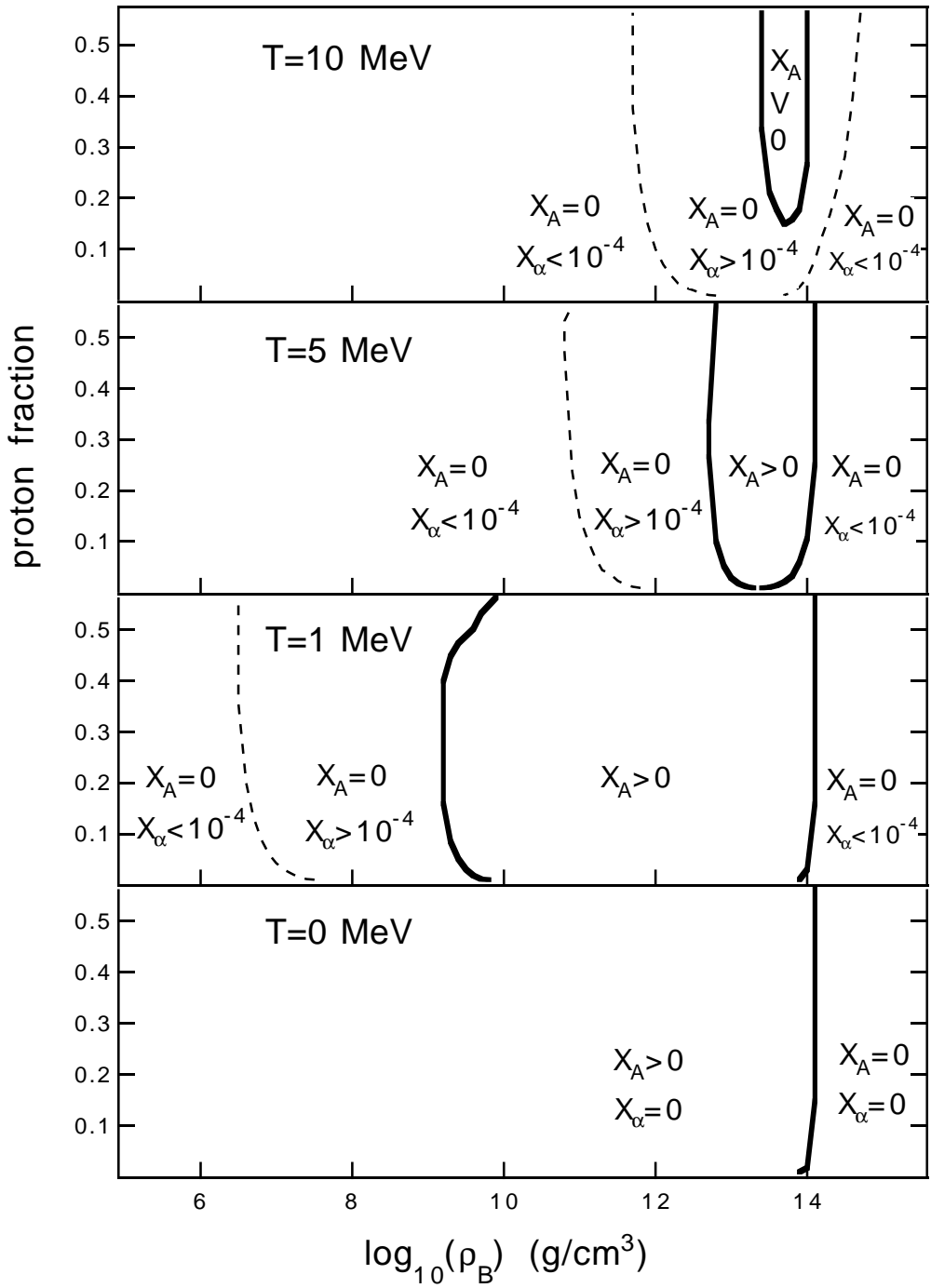
\item[Fig.3.] The phase diagrams of the nuclear matter at
$T=0, 1, 5,$ and $10 MeV$ in $\rho_B - Y_p$ plane.
The notations of $X_{\alpha}$, $X_A$ and the boundary curves are
the same as those in Fig.2.

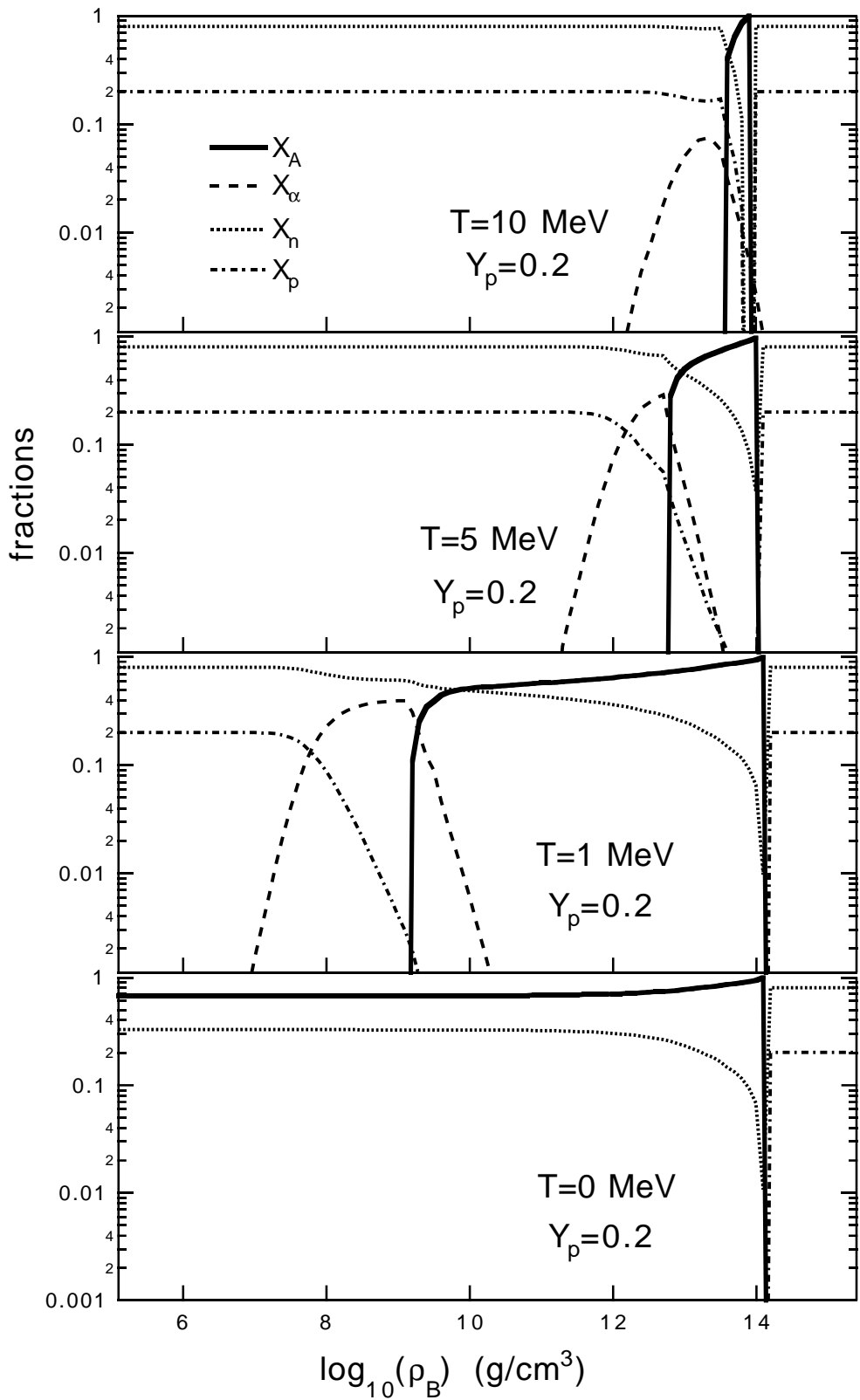
\item[Fig.4.] The neutron fraction $X_n$ (dotted curve), 
proton fraction $X_p$ (dot-dashed curve),
alpha-particle fraction $X_{\alpha}$ (dashed curve), 
and heavy nucleus fraction $X_A$ (solid curve)  
as functions of the baryon mass density $\rho_B$
at $Y_p=0.2$ with $T=0, 1, 5,$ and $10 MeV$.

\item[Fig.5a.] The nuclear mass number $A$ as a function of
the baryon mass density $\rho_B$ with the proton fraction
$Y_p=0.1, 0.2, 0.3, 0.4,$ and $0.5$ at $T=1 MeV$.

\item[Fig.5b.] The nuclear proton number $Z$ as a function of
the baryon mass density $\rho_B$ with the proton fraction
$Y_p=0.1, 0.2, 0.3, 0.4,$ and $0.5$ at $T=1 MeV$.

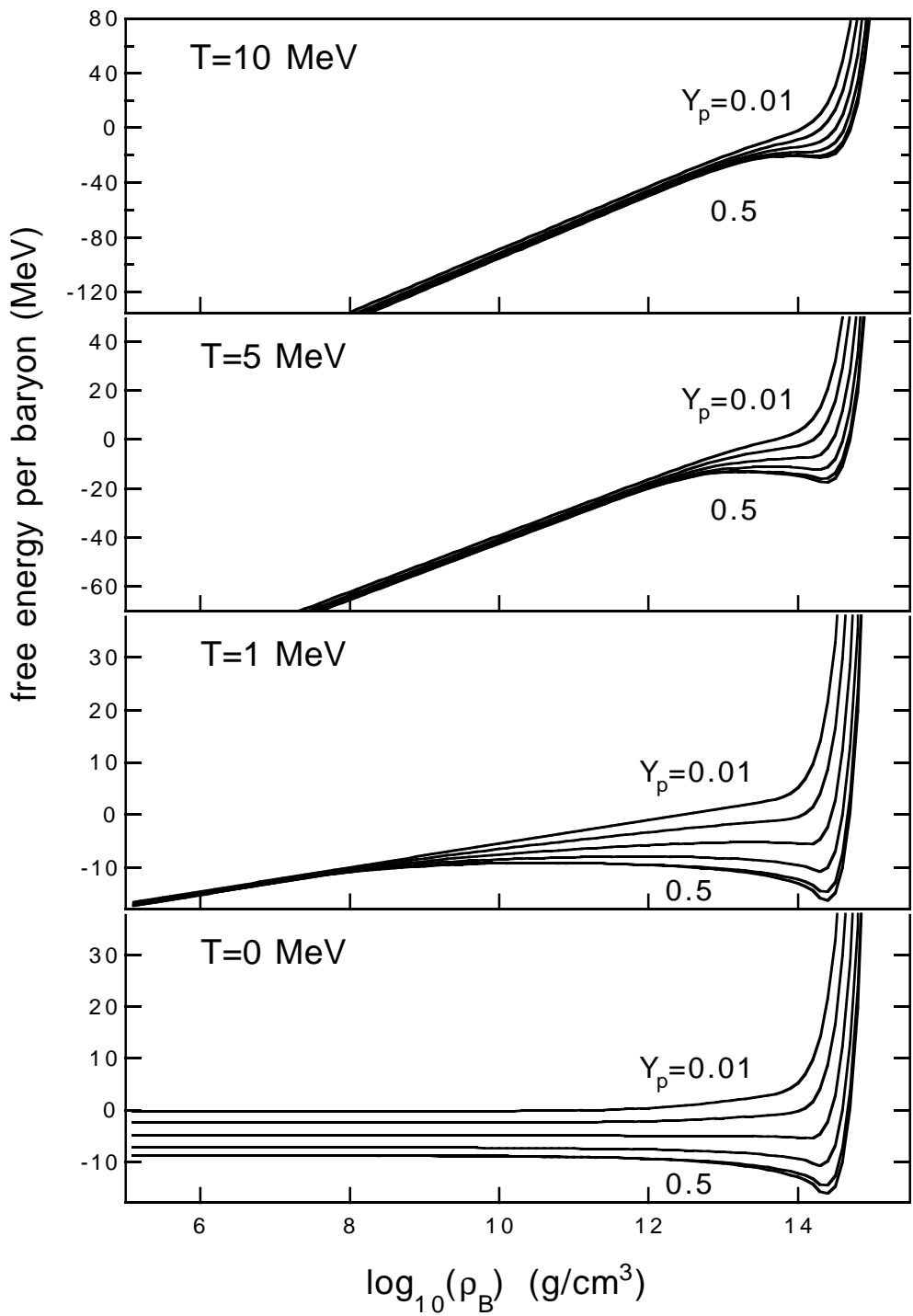
\item[Fig.6.] The free energy per baryon as a function of the baryon
mass density $\rho_B$ with the proton fraction 
$Y_p=0.01, 0.1, 0.2, 0.3, 0.4,$ and $0.5$
at $T=0, 1, 5,$ and $10 MeV$.

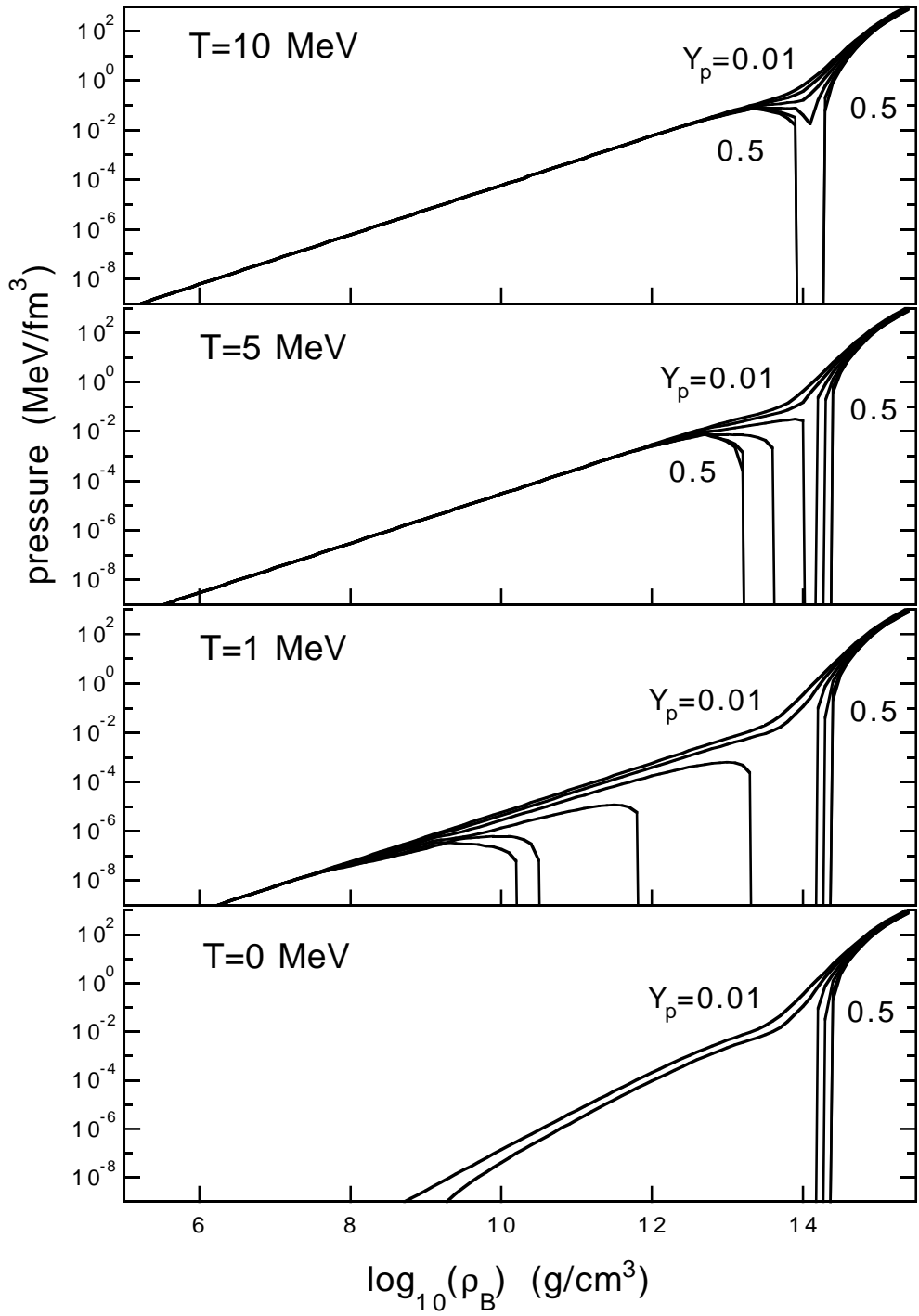
\item[Fig.7.] The pressure as a function of the baryon mass density
$\rho_B$ with the proton fraction $Y_p=0.01, 0.1, 0.2, 0.3, 0.4,$ and $0.5$
at $T=0, 1, 5,$ and $10 MeV$.

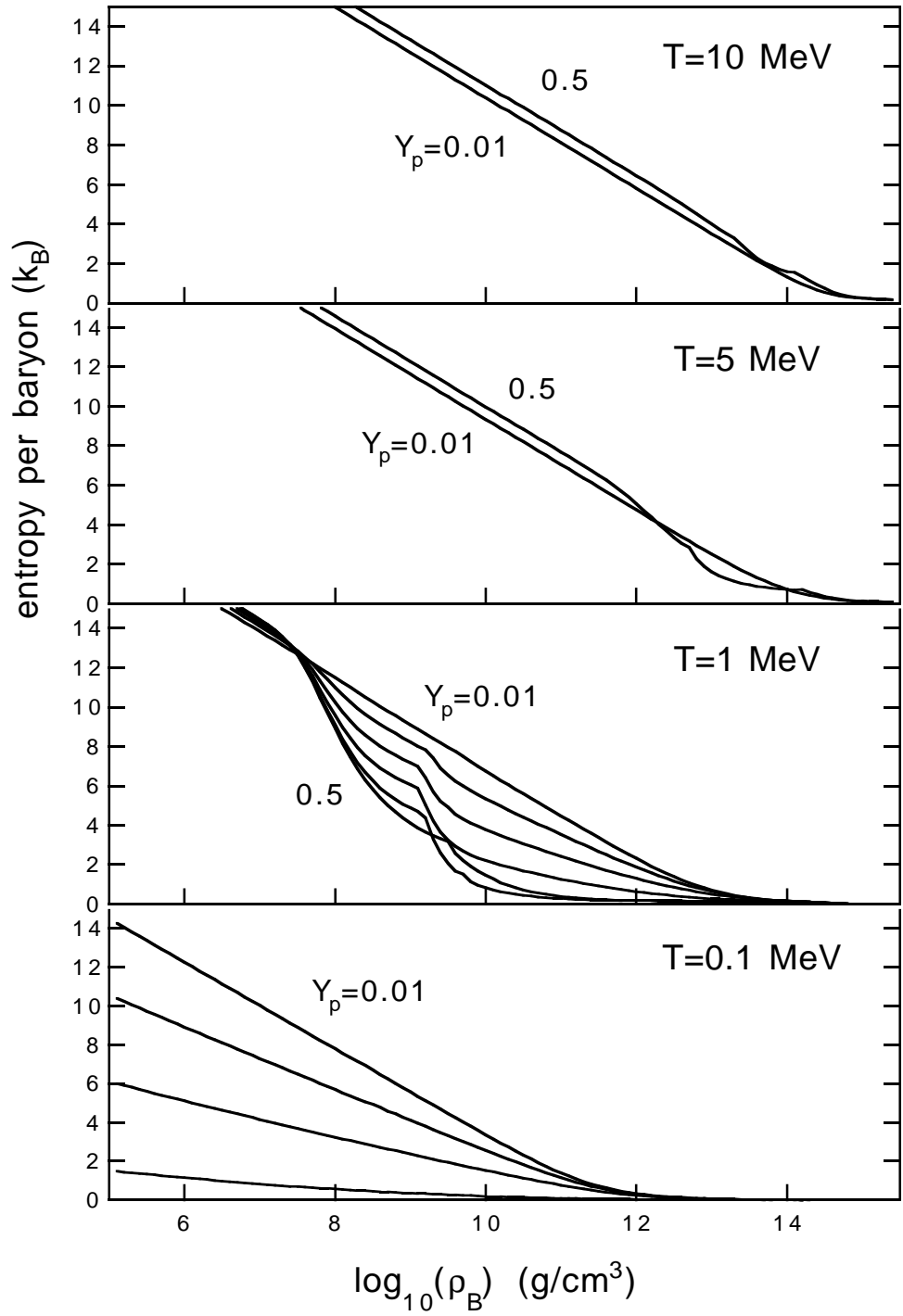
\item[Fig.8.] The entropy per baryon as a function of the baryon mass density
$\rho_B$ with the proton fraction $Y_p=0.01, 0.1, 0.2, 0.3, 0.4,$ and $0.5$
at $T=0.1$ and $1 MeV$, and with $Y_p=0.01$ and $0.5$ at $T=5$ and $10 MeV$.

\end{description}

\newpage


\begin{table}
    \caption{The parameter set TM1 for the lagrangian}
\vspace{0.5cm}
\begin{center}
\begin{tabular}{l l l r r r } 
\hline
\hline
 &     Parameter              & & &    TM1                  &  \\  \hline
 &   $M          \,\, [MeV]$  & & &    938.0                   &  \\
 &   $m_{\sigma} \,\, [MeV]$  & & &    511.19777                   &  \\
 &   $m_{\omega} \,\, [MeV]$  & & &    783.0                   &  \\
 &   $m_{\rho}   \,\, [MeV]$  & & &    770.0                   &  \\
 &   $g_{\sigma}$             & & &    10.02892                    &  \\
 &   $g_{\omega}$             & & &    12.61394                    &  \\
 &   $g_{\rho}  $             & & &     4.63219                    &  \\
 &   $g_{2} \,\, [fm^{-1}]$   & & &    -7.23247                    &  \\
 &   $g_{3}    $              & & &     0.61833                    &  \\
 &   $c_{3}    $              & & &    71.30747             &  \\  \hline
\end{tabular}
\end{center}
\end{table}

\begin{table}
    \caption{The ranges of the temperature $T$, 
             proton fraction $Y_p$, 
             and baryon mass density $\rho_B$ in the EOS table.
             The numbers of grid points in the ranges
             are also listed.  Note that the cases 
             of $T=0$ and $Y_p=0$ are covered with the
             listed ranges of the other parameters.}
\vspace{0.5cm}
\begin{center}
\begin{tabular}{l r r r } 
\hline
\hline
  Parameter             & Minimum  & Maximum & Grids   \\  
\hline
 $\log_{10}(T)\,\, [MeV]$           & -1.00     & 2.00     & 31      \\
 $\log_{10}(Y_p)$                   & -2.00     &-0.25     & 71      \\
 $\log_{10}(\rho_B)\,\, [g/cm^3]$   &  5.10     &15.40     & 104      \\
\hline
\end{tabular}
\end{center}
\end{table}


\begin{thebibliography}{999}

\bibitem{BM90} R.Brockmann and R.Machleidt, Phys. Rev. {\bf C42}, (1990) 1965.
\bibitem{BT92} R.Brockmann and H.Toki, Phys. Rev. Lett. {\bf 68 } (1992) 3408.
\bibitem{ST94} Y.Sugahara and H.Toki, Nucl. Phys. {\bf A579 } (1994) 557.
\bibitem{SST97} H.Shen, Y.Sugahara, and H.Toki, Phys. Rev.
                {\bf C55} (1997) 1211.
\bibitem{HS97} D.Hirata, K.Sumiyoshi, I.Tanihata, Y.Sugahara,
               T.Tachibana and H.Toki,
               Nucl. Phys. {\bf A616 } (1997) 438c.
\bibitem{HTT95} D.Hirata, H.Toki, and I.Tanihata,
               Nucl. Phys. {\bf A589 } (1995) 239.
\bibitem{HS96} D.Hirata, K.Sumiyoshi, B.V.Carlson, H.Toki, I.Tanihata,
               Nucl. Phys. {\bf A609 } (1996) 131.
\bibitem{MT97} Z.Y.Ma, H.Toki, B.Q.Chen, and N. Van Giai,
               Prog. Theor. Phys. {\bf 98} (1997) 917.
\bibitem{O93} K.Oyamatsu, Nucl. Phys. {\bf A561 } (1993) 431.
\bibitem{STOS98} H.Shen, H.Toki, K.Oyamatsu, and K.Sumiyoshi, 
                 Nucl. Phys. (1998) in press (nucl-th/9805035).
\bibitem{SW86} B.D. Serot and J.D. Walecka, Adv. Nucl. Phys. {\bf 16} (1986) 1.
\bibitem{LS91} J.M.Lattimer and F.D.Swesty, Nucl. Phys. {\bf A535 } (1991) 331.
\bibitem{SKT95} K.Sumiyoshi, H.Kuwabara, and H.Toki, Nucl. Phys. {\bf A581}
                (1995) 725.
\bibitem{cons} Physical Constants, Phys. Rev. {\bf D50} (1994) 1233.
\end{thebibliography}
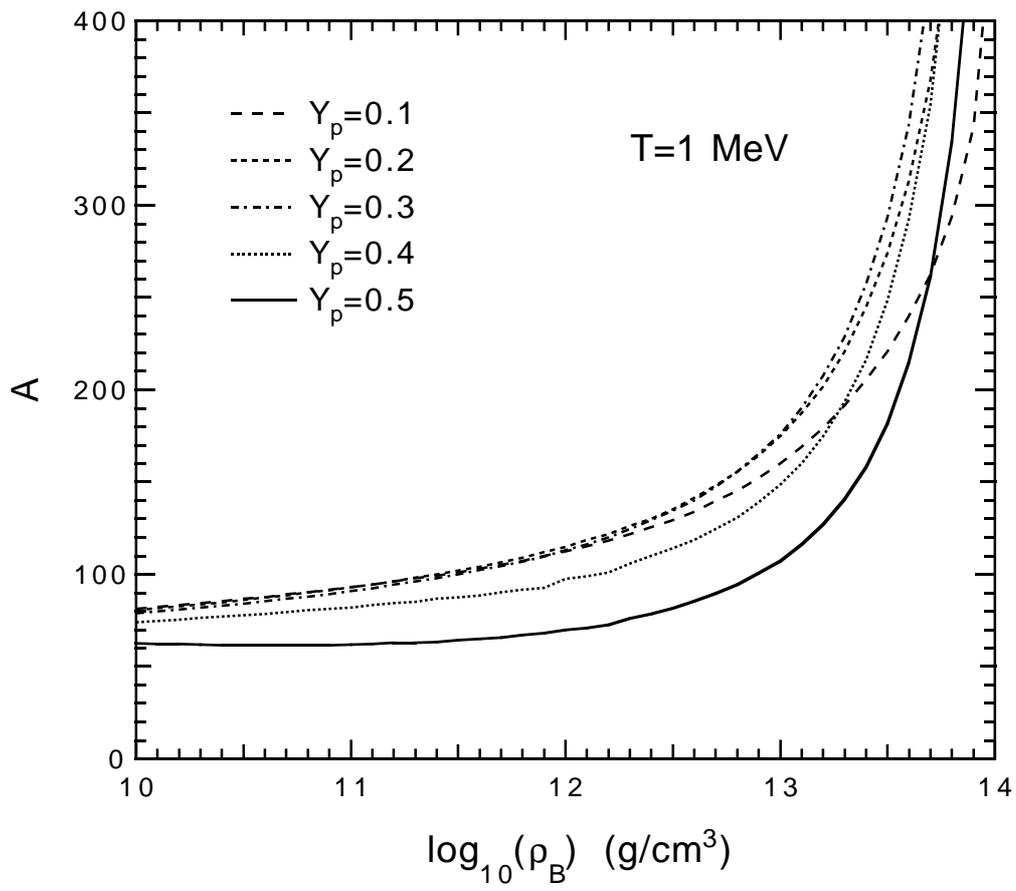
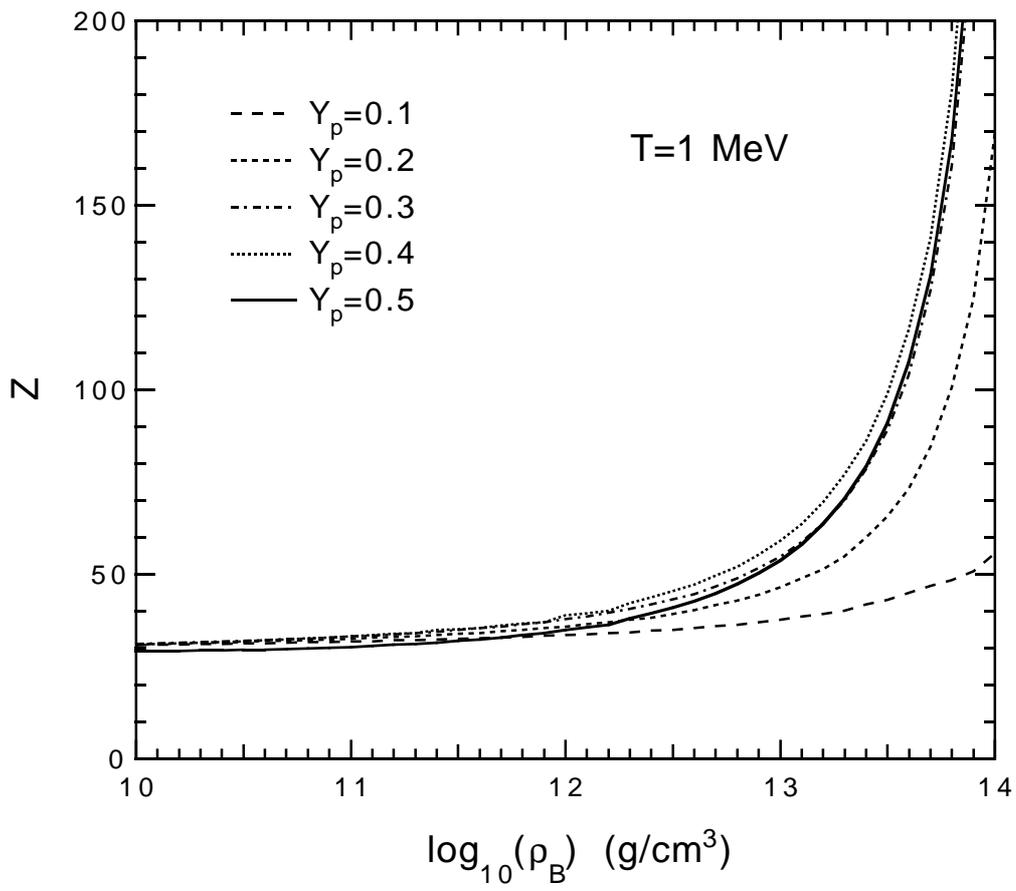
\end{document}